\documentclass[prl,nofootinbib,floats,floatfix,showpacs,twoside,preprint,superscriptaddress]{revtex4}

\usepackage{latexsym}
\usepackage{amsfonts}
\usepackage{amssymb}
\usepackage{amsmath}

\usepackage{graphicx}
\usepackage{amsbsy}
\usepackage{float}
\usepackage{epstopdf}
\DeclareGraphicsExtensions{.pdf,.eps,.png,.jpg,.mps}

\begin{document}
\title{Carbon dioxide emissions trading and hierarchical structure in worldwide finance and commodities markets}
\author{Zeyu Zheng}
\affiliation{Department of Environmental Sciences, 
Tokyo University of Information Sciences, Chiba 265-8501, Japan}
\author{Kazuko Yamasaki}

\affiliation{Department of Environmental Sciences, 
Tokyo University of Information Sciences, Chiba 265-8501, Japan}
\affiliation{Center for Polymer Studies and Department of Physics, Boston
University, Boston, MA 02215, USA}

\author{Joel Tenenbaum}
\affiliation{Center for Polymer Studies and Department of Physics, Boston
University, Boston, MA 02215, USA}

\author{H.~Eugene~Stanley}
\affiliation{Center for Polymer Studies and Department of Physics, Boston
University, Boston, MA 02215, USA}

\begin{abstract}
In a highly interdependent economic world, the nature of relationships between
financial entities is becoming an increasingly important 
area of study.  Recently, many studies
 have shown the usefulness of minimal spanning trees (MST) in extracting
interactions between financial entities.  Here, we propose 
a modified MST network whose metric distance is defined in terms of cross-correlation
coefficient absolute values, enabling the connections between anticorrelated entities to manifest
properly.  We investigate 69 daily time series, comprising 3 types of financial 
assets: 28 stock market indicators, 21 currency futures, and 20 commodity futures.
We show that though the resulting MST network evolves over time, the 
financial assets of similar type tend to have connections which are stable over time.
 In addition, we find a characteristic time lag between the volatility time series
of the stock market indicators and those of the EU CO2 emission allowance (EUA)
 and crude oil futures (WTI). This time lag is given by the peak of the cross-correlation function of the volatility time series EUA (or WTI) with that of the stock market indicators, and is markedly different ($>20$ days) from 0, showing that the volatility of stock market indicators today can 
predict the volatility of EU emissions allowances and of crude oil in the near future.
 
\end{abstract}

\maketitle

\section{Introduction and Method}
In the study of complex systems, there has been much work demonstrating the usefulness of extracting underlying structure from the correlations found in statistical data~\cite{Mantegna, Bonanno2, Bonanno1, Onnela1, Onnela2, Song, Micciche, Tumminello, ManandStan, Wang, laloux, Plerou1, Plerou2,Utsugi}. Any means of selection of statistically reliable information from correlation matrices
has been dubbed a ``filtering procedure''~\cite{Tumminello}. Useful examples of filtering procedures 
that use a correlation matrix from return time series are: hierarchical clustering~\cite{Mantegna, Bonanno2, Bonanno1, Onnela1, Onnela2, Song,Micciche}, procedures based on the random matrix theory~\cite{Wang, laloux, Plerou1, Plerou2,Utsugi}, and networks from minimum spanning trees~\cite{Mantegna, Bonanno2, Bonanno1, Onnela1, Onnela2}.

Correlation structure studies are not limited only to stock return time series~\cite{Mantegna, Bonanno2, Bonanno1},
but also extend to quasi-synchronously recorded time series of worldwide stock exchange market indices~\cite{Bonanno1} and stock return volatility increments~\cite{Micciche}.

Financial time series can include not only stock price, but also
many other types of data, such as commodities price, treasury yield,
market index, etc....  Investigation 
of multi-type quasi-synchronous financial data may
yield insights into the interdependent relationships
of markets and commodities.  Moreover, a relationship map of financial
assets can highlight the movement of speculative capital.

We investigate 69 daily financial series from the time period spanning January 2007 to September 2011. The data
set includes 21 currency futures, 20 commodity futures which are taken from {\tt http://data.theice.com}, and 28 stock market indicators, which are taken from {\tt http://finance.yahoo.com} (see
Appendix).

Recently, papers have shown the usefulness of the correlation structure described by an ultrametric space and a corresponding hierarchical organization for financial return time series~\cite{Mantegna, Bonanno2, Bonanno1}. The approach requires the definition of a metric distance. 
Because correlation does not fulfill the three~\cite{metric} axioms that define a metric,
the aforementioned papers use the Mantegna-Sornette distance defined by

\begin{equation}
  d_{ij}=\sqrt{2(1-\rho_{ij})}
  \label{eq5}
\end{equation}

\noindent for each pair of elements $i$ and $j$, where $\rho_{ij}$ is the correlation coefficient of the two time series given by~\cite{feller}

\begin{equation}
  \rho_{ij} \equiv \frac{\langle Y_iY_j\rangle-\langle Y_i\rangle \langle Y_j\rangle}{\sqrt{(\langle{Y_i}^2\rangle-\langle Y_i\rangle^2)(\langle{Y_j}^2\rangle-\langle Y_j\rangle^2)}} = \frac{\Big\langle \bigg(Y_i-\langle Y_i \rangle\bigg)\bigg(Y_j-\langle Y_j \rangle\bigg) \Big\rangle}{\sigma_{Y_i}\sigma_{Y_j}},
  \label{eq6.5}
\end{equation}

\noindent where $\langle ... \rangle$ denotes the mean.

This distance $d_{ij}$ fulfills the three axioms~\cite{metric} of a metric: i) $d_{ij}=0$ if and only if $i=j$, ii) $d_{ij}=d_{ji}$, and iii) $d_{ij} \leq d_{ik}+d_{kj}$ ~\cite{Mantegna,Bonanno1,Bonanno2}. 

In this work, we make a modification to the metric above, based on the reasoning that if two time series have a very large {\it negative} correlation, they should still be considered close to each other in ultrametric ``correlation'' space, since this would indicate a strong connection, regardless of the sign.  Likewise, weak correlations of either sign indicate a weaker connection.
On this basis, strong correlations of either sign should be considered closer than weak ones.  

The value of this modification can be readily seen in situations where two entities have strong { \it anti}-correlations, as has been observed in the relationship between bond and stock markets (e.g. between UK gilts and the FTSEMIB), between stocks and currency futures, and  between industries and their inputs (e.g. the price of oil and the value of airline stock)~\cite{Norden, Kwan,Hammoudeh,businessweek}. Here, use of the conventional Mantenga-Sornette metric would likely result in any of these two entities manifesting on opposite sides of a tree, even though we know them to be very closely linked. Use of our
generalized metric ensures that such entities will be placed nearby when such appropriately strong relations exist.

For this reason, we replace the Pearson correlation coefficient in Eq.~\ref{eq5} with the absolute
value of the Pearson correlation coefficient, defining a new distance as
\begin{equation}
    d_{ij}=\sqrt{2(1-|\rho_{ij}|)}
 \label{eq6}
 \end{equation}

\noindent with the $\rho_{ij}$ defined as before in Eq.~\ref{eq6.5} as the correlation coefficient of assets $i$ and $j$.

This equation also fulfills three axioms of a metric distance. Because the set of assets considered has no cases such that $\rho_{ij}=-1$, the first axiom is satisfied on this set in that  $d_{ij}=0$ if and only if
the correlation is total ($\rho=1$, meaning that the stocks perform the same stochastic process). The second axiom 
(that of symmetry) is trivially satisfied because $\rho_{ij}=\rho_{ji}$ by definition of the
Pearson correlation coefficient.
For the validity of axiom (iii), consider three time series ${Y_i}$, ${Y_j}$ and ${Y_k}$, which have means
equal to 0 and standard deviations equal to 1.  All times series have the same length.
In order to prove $d_{ij} \leq d_{ik}+d_{kj}$, firstly we define two new time series as

\begin{equation}
 Y'_i\equiv \begin{cases}
 Y_i,&\mbox{if }  \rho_{Y_i, Y_k} \geq 0 \\
 -Y_i,&\mbox{if } \rho_{Y_i, Y_k} < 0)
\end{cases}
\end{equation}
\begin{equation}
 Y'_j\equiv \begin{cases}
 Y_j,&\mbox{if }  \rho_{Y_j, Y_k} \geq 0 \\
 -Y_j,&\mbox{if } \rho_{Y_j, Y_k} < 0.
\end{cases}
\end{equation}

\noindent So we can rewrite $d_{ik}+d_{kj}$ of distance Eq.~\ref{eq6} by using $Y'_i$ and $Y'_j$  as
\begin{eqnarray}
d_{ik}+d_{kj} & =\sqrt{2(1-|\rho_{Y_i,Y_k}|)}+\sqrt{2(1-|\rho_{Y_j,Y_k}|)}&\\
&  = \sqrt{2(1-\rho_{Y'_i,Y_k})}+\sqrt{2(1-\rho_{Y'_j,Y_k})}\\
&  \geq \sqrt{2(1-\rho_{Y'_i,Y'_j})}\\
&\begin{cases}
 =\sqrt{2(1-|\rho_{Y'_i,Y'_j}|)}, &\mbox{if } \rho_{Y'_i, Y'_j} \geq 0 \\
 \geq \sqrt{2(1-|\rho_{Y'_i,Y'_j}|)}, &\mbox{if } \rho_{Y'_i, Y'_j} < 0 \\
\end{cases}\\
& =\sqrt{2(1-|\rho_{Y_i,Y_j}|)} \equiv d_{ij}. &
\end{eqnarray}

\noindent Thus $d_{ik}+d_{kj} \geq d_{ij}$ , and our metric satisfies the three axioms of a metric.

For each of the 22 financial time series, we calculate the return time series, defined as the change of logarithmic price of time series $i$
\begin{equation}
 R_i(t) \equiv \ln(Y_i(t+1))-\ln(Y_i(t)).
\label{eq2}
\end{equation}
Here $Y_i(t)$ is the daily price time series of financial asset $i$.
For each of the 22 time series, we also calculate the volatility time series which is defined simply as the
absolute value of the return $|R_i|$
\begin{equation}
 V_i(t) \equiv |R_i(t)| = |\ln(Y_i(t+1))-\ln(Y_i(t))|.
\label{eq3}
\end {equation}

Additionally, we define the cross-correlation for our analysis.  Consider two time series $\{y_t\}$ and $\{y'_t\}$.  The cross-correlation between  $\{y_t\}$ and $\{y'_t\}$ is given by
\begin{equation}
 C_{y,y'}(n)\equiv \overline{(y_t-\mu)(y'_{t+n}-\mu'))}/(\sigma\sigma'),
\label{eq1}
\end{equation}
where $\mu$ and $\mu'$ are the respective means and $\sigma$ and $\sigma'$ are the respective standard deviations of the series $\{y_i\}$ and $\{y_i'\}$.

The efficient market hypothesis, a basic tenet of modern economics, states that markets are approximately efficient, meaning that one cannot consistently achieve returns better than the market because all information about
an asset is already incorporated into that asset's price~\cite{Samuelson,ManandStan}. As a result, it is believed that the long-range memory cannot exist in any return time series. Suppose that long-range auto-correlations
 exist in a return time series: investors may then obtain benefits by using information, which stands in contradiction to the principle of an efficient market.
Consider the cross-correlation function (Eq. (\ref{eq1})) between  return time series of asset $i$ and asset $j$. Any significant cross-correlations $C_{R_i,R_j}(n)$ in  $n \neq 0$ of two return time series would also contradict the
 existence of an efficient market. Therefore we can assume that significant cross-correlations $C_{R_i,R_j}(n)$ will only exist for the case $n=0$. 

However, because trading occurs at different times in different cities,
some markets are open when others are closed.
The effect of non-synchronous trading in time 
series analysis has been well stated~\cite{Lo, Lin}. In fact,
the highest degree of correlation between different markets may be 
detected at a one-day time lag because of the time difference. 

Therefore, significant cross-correlations $C_{R_i,R_j}(n)$ may also exist for $n=-1$ or $n=1$. Additionally, we only care about the magnitude and not the sign of cross-correlation. Thus, we define the absolute correlation coefficient as 
\begin{equation}
\rho_{ij}=\max \left (\left \vert C_{R_i,R_j}(n)\right \vert \right )
\label{eq4}
\end {equation}
for $n=-1, 0, 1$.

In volatility time series, long-range correlations $C_{V_i,V_j}(n)$ have been shown to exist [10-13].  It follows that significant cross-correlations $C_{V_i,V_j}(n)$ may exist for $n\gg0$ or $n\ll0$. Additionally, the existence of long-range negative correlations between past returns and future volatility \cite{Bouchaud,Perello}, known as the leverage effect, has also been reported. This correlation is moderate and decays exponentially over the long term. However, while both of these types of correlations may help predict the financial risks on a long-range time interval, we point out that neither the negative correlation between returns and volatilities nor long-range autocorrelation of volatility can be used to obtain benefits. This is because the price volatility does not include the direction of price changes, and so neither contradicts the efficient market hypothesis.  

\begin{figure}[b]
\centering \includegraphics[width=0.9\textwidth]{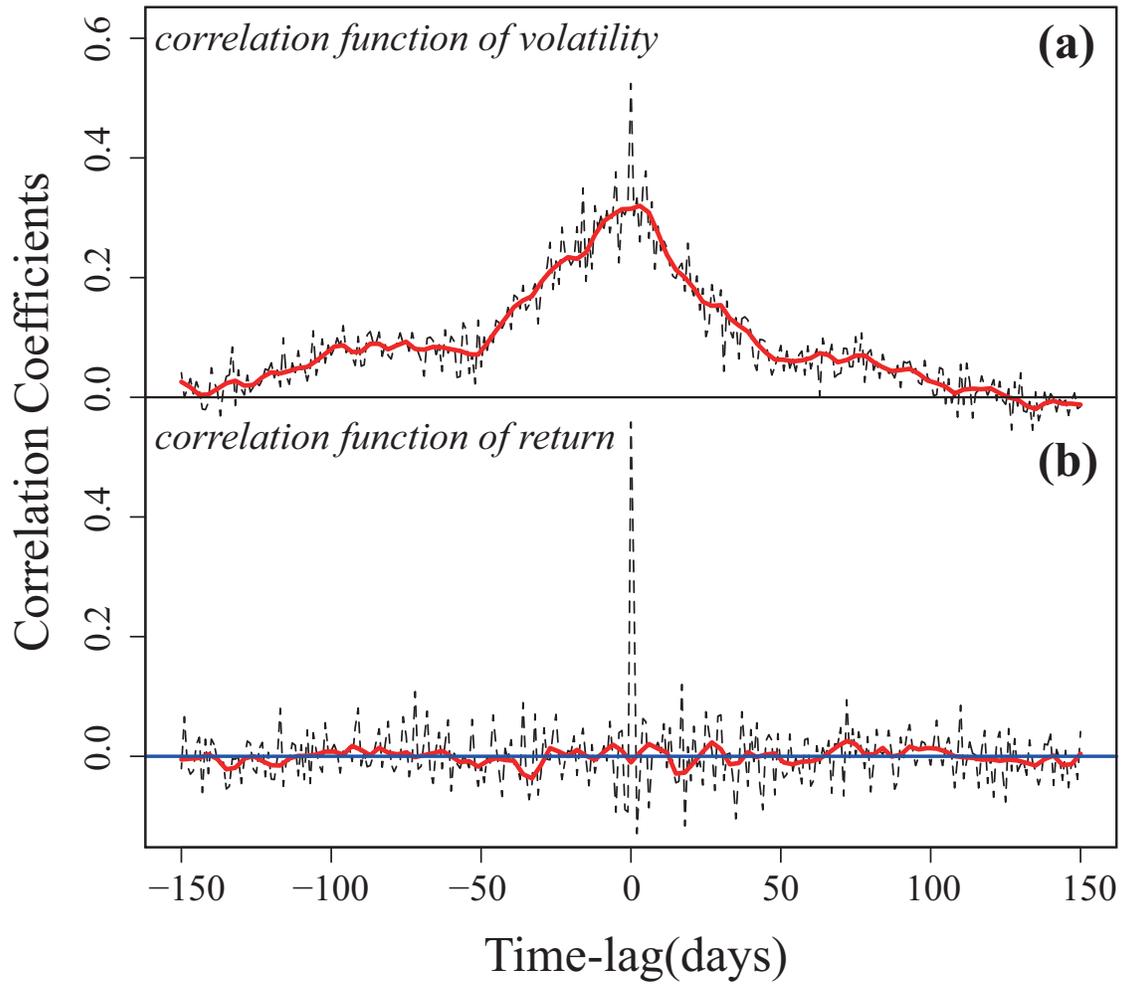}\\
\caption{ The cross-correlation function $C(n)$ of volatility (a) and return (b) time series between the DJIA and FTSE100. Both show statistically significant correlation coefficients 
at their maxima near time lag=0 (dotted curve). Solid lines show the LOWESS (locally weighted scatter plot smoothing) values of $C(n)$, smoothed over a span of 30 days. 
}
\label{1}
\end{figure}

As a sample, the correlation functions of volatility $C_{V_i, V_j}(n)$ and return $C_{R_i, R_j}(n)$ between the FTSE100 and DJIA are shown in Fig.~\ref{1}. The peaks (highest correlations) of both correlation functions are near n=0; however, the correlation function for returns
is fast decaying, quickly approaching 0 for $n \neq 0$, while the volatility correlation function 
is slow decaying with $C_{V_i, V_j}(n) > 0$ for $n > -50$ and $n < 50$. 

We now turn our attention to the stability and structure of MSTs, which are made using the distance based defined in Eq.~\ref{eq6}.  Following our discussion on MSTs, we will show the correlation function graphs of volatility that relate the correlation $C(n)$ to the time-lag $n$, specifically focusing on the value of $n$ that gives
the LOWESS (the smoother which uses locally weighted polynomial regression)  equal to its maximum value.
Here, the LOWESS is defined by a complex algorithm, proposed by W. S Cleveland \cite{Cleveland} in 1981. For each value, we define 10th nearest neighbor values as the local region, which is used to calculate the LOWESS value.

An MST is  defined as the set of $n-1$ links that connect a set of $n$ elements 
in the smallest possible total distance~\cite{Graham}. MSTs have been used in prior papers~\cite{Mantegna, Bonanno2, Bonanno1, Onnela1, Onnela2} to connect financial data, illustrating the MST's usefulness in highlighting the interactions 
between a number of financial time series. 

\begin{figure}[b]
\centering \includegraphics[width=0.99\textwidth]{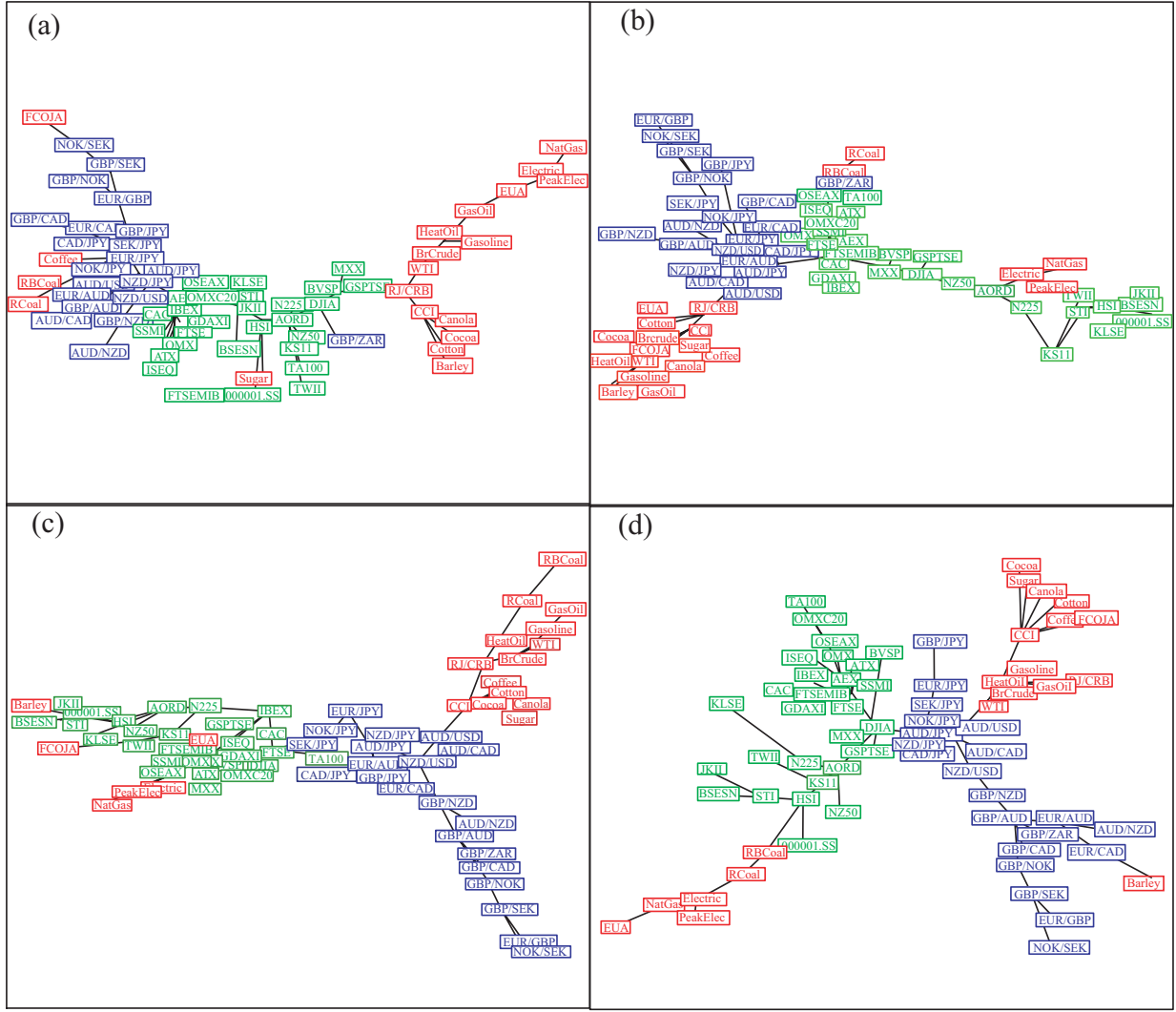}\\
\caption{The MST  obtained  from the absolute correlation coefficients $|\rho_{ij}|$ of the set of 68 return time series during in individual calendar years (a) 2007 (b) 2008 (c) 2009 (d) 2010. Red indicates commodity futures, blue indicates currency futures, and green represents stock market indicators.  See Appendix for a listing of symbols and their meanings.
}
\label{2}
\end{figure}

In Fig.~\ref{2}, we find that, although MSTs show significantly different structures in different calendar years,
the same type of financial assets tend to group together consistently over time.

\begin{figure}[b]
\centering \includegraphics[width=0.9\textwidth]{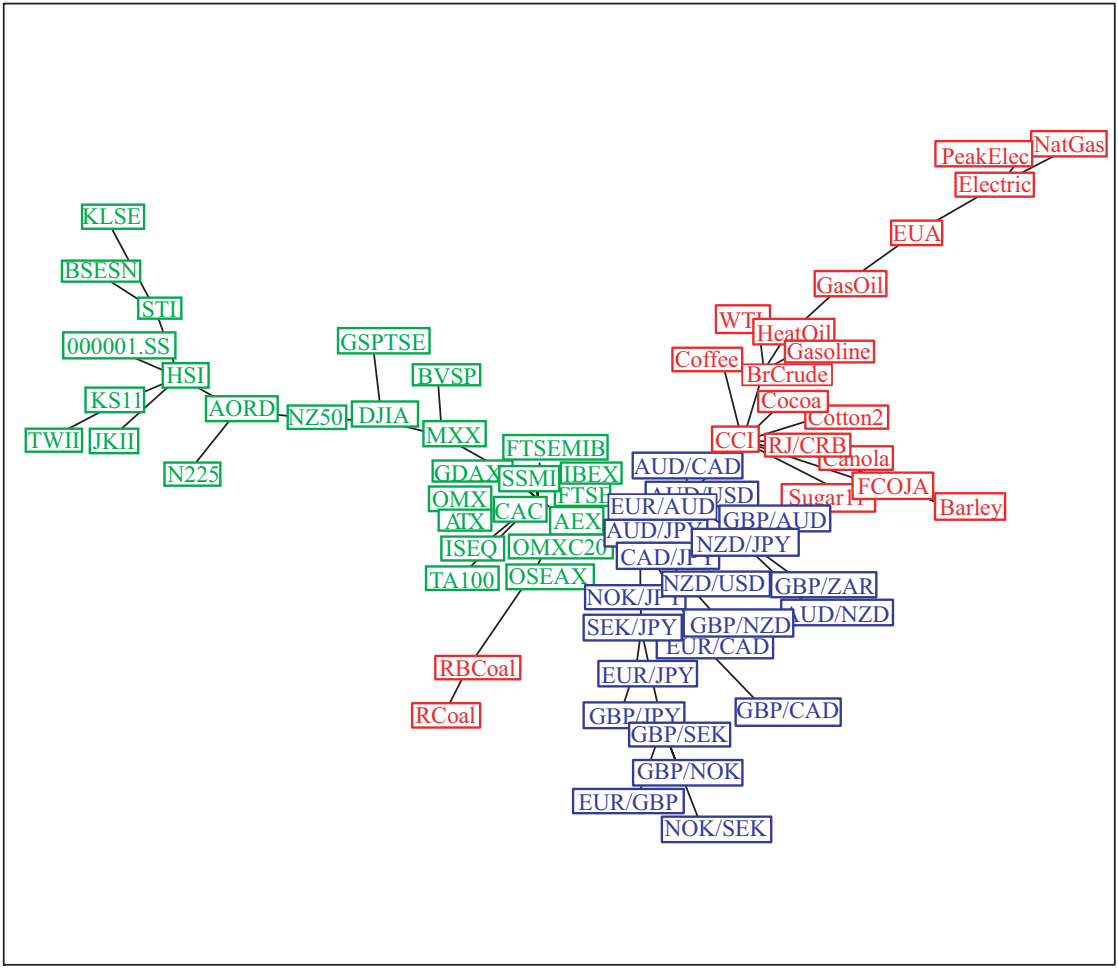}\\
\caption{The minimal-spanning tree (MST), similar to Fig.~\ref{2}, but for the longer time span January 2007 to September 2011. See Appendix for symbol definitions.}
\label{3}
\end{figure}

We also find that the stock market indicators (blue in Fig.~\ref{2}) and currency futures (green) groups show stronger interconnections than the commodities (red) group. For stock market indicators and currency futures, financial factors
are the predominant reason for price changes.  On the other hand, commodity futures may be just as much affected by investment 
as they are by actual supply and demand.  Speculation in commodity futures may alter pricing 
in a way that contradicts the law of supply and demand.
The existence of such contradictions depends on commodity type and the calendar year and therefore serve to decrease the stability of the MST.  If this reasoning is correct, increasing the time span of the time series for cross-correlation should make the 
observed connections more stable.
In Fig.~\ref{3}, we show the MST using the time series from January 2007 to September 2011.  We note
that only two coal futures are not connected to the commodity group and the stability is greater
than for single-year time series.

\begin{figure}[b]
\centering \includegraphics[width=0.9\textwidth]{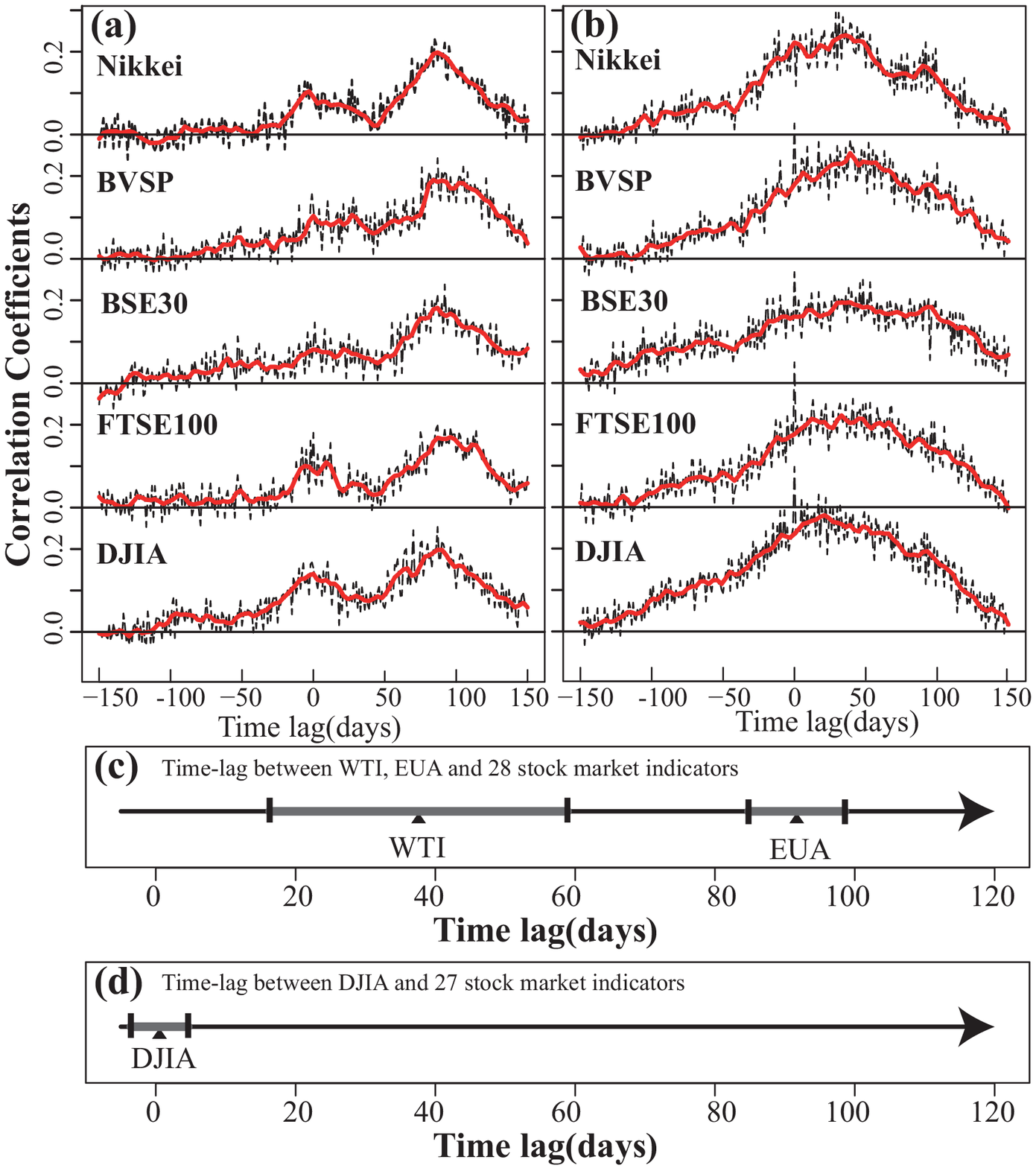}\\
\caption{Cross-correlation function $C(n)$ of volatility daily time series between 5 main stock
market indices and (a) EUA (European carbon emissions permits), or (b) WTI (light sweet crude oil). Red lines indicate the locally weighted scatter plot smoothing values (LOWESS) of $C(n)$. Graphs show  systemic time shift for the highest cross-correlation value. This time shift is observed in most stock markets.
(c) indicates the average time-lag between EUA, WTI, and 28 stock market indicators, with error bars showing the standard deviations. (d) indicates the average time-lag between the DJIA and other 27 stock market indicators, with error bars showing the standard deviations. The time lags (in days) are calculated from the time lag $n$ of highest LOWESS values.}
\label{4}
\end{figure}

In Fig.\ref{4}, we describe the cross-correlation functions of the volatility time series. We show the cross-correlation function of main stock market indicators with EUA in (a), and with WTI in (b). A systemic time shift between EUA or WTI, and stock market indicators can clearly be seen.  Since the maximum cross-correlation coefficient in most 
functions is not much greater than 0.2, the connections are not so strong, but certainly they are significant.
Further tests of Granger causality also show that the volatility of stock indices is useful in forecasting
the volatility of EUA and WTI approximately 20 to 120 days in advance.

As mentioned before, the correlation function of volatility is a slow-decaying function. It is much more slowly decaying than the correlation function of return time series (see Fig.~\ref{1}), meaning that a long-range cross-correlation relationship exists.  If we consider significant cross-correlations between different volatility time series 
to be an information transfer between different assets, the time lag corresponding to the highest 
values of correlation gives the time lag of that information transfer.  It is worth pointing out that the time lag between each pair of stock markets is approximately 0, such as the time lag between DJIA and other 27 stocks indicators are shown in Fig.~\ref{4} (d) . 
We show a simple summary of such time lag in Fig.~\ref{4} (c) and (d) If there
is information affecting stock markets at a time of 0 days, the EUA and WTI crude oil futures will be affected by this information roughly 30 and 90 days later respectively.  Because both crude oil and the ability to emit carbon are major inputs in the world economy, the existence of this time lag can have strong implications in terms of potential economic feedback loops.

\section{Discussion}
The drawn MSTs are reflective of a number of easily reasoned underlying economic relationships, both through the stability and specificity of the links. 

As expected, we find a stable tendency for like financial assets to cluster.  Even during the otherwise anomalous 2007 subprime 
lending crisis and 2008 global financial crisis, this clustering tendency is preserved.
This indicates the existence of strong stable connections, which come out of the strong cross-correlations, reflective of basic economic features and interactions.  These connections are stable over time and 
not affected by market conditions.

On the other hand, certain portions of the MSTs are consistently unstable, like those relating to coal. This also may be reflective of economic relationships.  Unlike other commodities like oil, 
speculation in coal is limited, so the movement of the coal futures may be simple supply and demand, 
as opposed to driven by speculation.  Coal's lack of strong connections to other commodities 
may be a result of investor's low speculation in stock when building their commodities portfolios.

Certain connections in our MSTs reveal underlying relationships that are created by regulation.  For example, we find that EUA  futures mostly connect with the base electricity and natural gas futures, which show stable correlations among them. EU allowance permits, as a part of the European Union Emission Trading Scheme, are either allocated or auctioned and allow
a firm to emit a designated amount of carbon dioxide~\cite{Stavins}.  Since power generation accounts for about one-quarter of total emissions of carbon dioxide, and natural gas is the most resource of electricity generation in UK, the stable connections of EUA to UK base electricity futures and UK natural gas futures in our MST graph is reasonable. 

 Also intuitive is the location specific clustering of stock market indices.  AORD, an Australian index, consistently
appears closely linked to those of nearby countries like New Zealand, Japan, and China.  Similarly, the HSI not only keeps
 connections with most of the Asian stock indices like the JK11, 000001.SS, BSESN, and TWII, but also keeps connections with or otherwise stays closely connected to indices from 
America and Australia.  Thus, the MST created reflects the common knowledge
that Hong Kong is the financial center of Asia.

One benefit of the novelty of our approach is that it connects indices of dissimilar type, yielding new insights.  The index that connects most to coal is OSEAX, that of Norway.   Norway is rich in oil and natural gas, which explains why the Norway stock index appears as the most ``coal-like'' of the national indices.  Similarly, the most ``currency-like''	 of the national stock indices seems to be AEX, the Dutch securities index. 

We also note the relationship between the centrality in the network and real world geographical knowledge.  SOK/SEK
is a currency future that is among the furthest from the center.  This is intuitive, since currency trading between Norway and Sweden has little to do with financial activity in the rest of the world.  The same principle applies to the trading of
Euros with the British pound, shown as EUR/GBP.  The Australian dollar, on the other hand, plays a central role in its exchanges with far away currencies like the US dollar, Euro, Japanese yen, and Canadian dollar.

\section{Conclusion}
In this paper, we have analyzed the correlation function of return and volatility time series, constructed MSTs
based on return  time series, and found consistent time lags in the correlation functions 
of the volatilities.  From these analysis, we have two main conclusions. (i) The stability of MST structure clustering between like commodities reflects a basic rule of economic activity, that the interaction between
economic actors is not easily affected by capital movement. The method of absolute cross-correlation coefficient based MSTs has strong implications in the ongoing debate about the relationships of different financial commodity time series. (ii) We find that a  time lag of correlation functions of volatility appears between stock markets and EUA and WTI.
From this finding, we hypothesize that there may be systemic differences in the spread of financial risk, most often quantified as volatility. In other words, as concerns risk, different types of markets may have different sensitivities to economic information and other influencing factors.  It would be interesting, from theoretical point of view, to generalize this time lag  to predict the financial risks on a much longer time interval. However, much more would need to be understood first, such as the properties and mechanisms for this time lag.  Hence, further empirical study is needed first.  We endeavor to address this question more in future work.

\section{Appendix: Set of Data}
The data under investigation includes 28 stock market indicators, 21 currency futures, and 20 commodity futures.

The stock market indicators investigated are:
\begin{center} {\scriptsize
  \begin{tabular}{|c|l|l||c|l|l|}
    \hline
    Symbol & Meaning & Notes & Symbol & Meaning & Notes  \\

    \hline
    000001.SS & SSE Composite Index & Shanghai stocks & ISEQ & Irish Stock Exchange Quotient &\\      
    AEX & Amsterdam Exchange Index & Dutch securities &  JKII & Jakarta Islamic index &\\
    AORD & All Ordinaries          & Australian stocks &  KLSE & Kuala Lumpur Stock Exchange &\\
    ATX & Austrian Traded Index    &                   & KS11 & Seoul Composite (South Korea) &\\
    BSESN & SENSEX                 & Bombay stocks     & MXX & Mexican Stock Exchange IPC &\\ 
    BVSP & Bovespa Index           & S\~ao Paulo stocks & N225 & NIKKEI 225 & Tokyo stocks\\ 
    CAC & CAC 40                   & French stocks      &  NZ50 & NZX 50 Index & New Zealand index\\    
    DJIA & Dow Jones Industrial Average & American index&  OMX & OMX Stockholm 30 & \\
    FTSE & FTSE 100                & London stocks  &  OMXC20 & OMX Copenhagen 20 & \\    
    FTSEMIB & FTSE MIB             & Italian stocks & OSEAX & Oslo B\o rs All Share Index &\\
    GDAXI & Deutscher Aktien Index & German blue chips & STI & Straits Times Index & Singapore stocks\\ 
    GSPTSE & S\&P/TSX Composite Index & Toronto stocks & SSMI & Swiss Market Index & \\ 
    HSI & Hang Seng Index          & Hong Kong stocks & TA100 & Tel Aviv 100 & \\ 
    IBEX & IBEX 35                 & Spanish stocks & TWII & TSEC weighted index &Taiwanese stocks\\ 

    \hline
  \end{tabular}}
\end{center}

The commodity and currency futures investigated are all traded in the markets of intercontinental exchange. The commodity futures are:

\begin{center} {\scriptsize
  \begin{tabular}{|c|l|l||c|l|l|}
    \hline
    Symbol & Meaning & Notes & Symbol & Meaning & Notes  \\
    \hline
    Barley &  Western Barley Futures &            ~        & Gasoline &  RBOB Gasoline Futures  &\\
    BrCrude & Brent Crude Futures    & North Sea crude oil \hfill& HeatOil & Heating Oil Futures  &\\
    Canola & Canola Futures          &             ~          & NatGas & UK Natural Gas Futures  &\\
    CCI & Consumer Confidence Index Futures &              & PeakElec & UK Peak Electricity Futures  &\\
    Cocoa & Cocoa Futures &                      ~          & RBCoal & Richards Bay Coal Futures  &\\
    Cotton & Cotton No. 2 Futures &              ~         & RCoal & Rotterdam Coal Futures  &\\
    FCOJA & FCOJ-A Futures & Florida orange juice          & RJ/CBR & Thomson Reuters/Jefferies CRB Index&assorted commodities \\
    Electric & UK Base Electricity Futures &       ~       & Sugar & Sugar No. 11 Futures  &\\
    Coffee & Coffee ``C'' Futures &                 ~     & WTI & West Texas Intermediate & Texas crude oil\\
    GasOil & Gas Oil Futures &                  ~         & EUA & EU emission allowance &\\
    \hline
  \end{tabular}}
\end{center}

Additionally, the currency futures in the form A/B refers to the value of currency A in units of currency B.  For
example, USD/EUR would be the value of US dollars in units of Euros.  The currency futures investigated are:
\begin{center} {\footnotesize
  \begin{tabular}{|c|l|l|}
    \hline
    Symbol & Currency & Traded in units of\\
    \hline
    AUD & Australian dollar & CAD, JPY, NZD, USD\\
    CAD & Canadian dollar   & JPY\\
    EUR & Euro              & AUD, CAD, GBP, JPY\\
    GDP & British pound     & AUD, CAD, JPY, NOK, NZD, ZAR, SEK\\
    JPY & Japanese yen      & \\
    NOK & Norwegian krone   &JPY, SEK\\
    NZD & New Zealand dollar &JPY, USD\\
    SEK & Swedish krona     & JPY\\
    ZAR & South African rand & \\
  \hline
  \end{tabular}}
\end{center}


\begin{thebibliography}{99}

\bibitem{Mantegna}
R. N. Mantegna,
Eur. Phys. J. B {\bf11}, 193-197 (1999).
\bibitem{Bonanno1} 
G. Bonanno, N. Vandewalle  and  R.N. Mantegna,
 Phys. Rev. E {\bf62}, R7615-R7618 (2000).
\bibitem{Bonanno2} 
 G. Bonanno, F. Lillo, and  R.N. Mantegna,
 Quantitative Finance {\bf1}, 96-104 (2001).
\bibitem{Onnela1} 
J.-P. Onnela, K. Kaski and J. Kertesz,
Eur. Phys. J. B {\bf38}, 353-362 (2004).
\bibitem{Onnela2}
J.-P. Onnela, A. Chakraborti, K. Kaski, J. Kertesz, and A. Kanto,
Phys. Rev. E {\bf 68}, 056110 (2003)
\bibitem{Song}
D. Song, M. Tumminello, W. Zhou, and R. N. Mantegna,
Phys. Rev. E {\bf 84}, 026108 (2011)
\bibitem{Micciche}
S. Miccichea, G. Bonannob, F. Lilloa, and R. N. Mantegna,
Physica A  {\bf324}, 66-73, (2003)
\bibitem{Tumminello}
  M. Tumminello, F. Lillo, and R. N. Mantegna, 
  J. Econ. Behav. Org. {\bf75} 40 (2010); Phys. Rev. E, {\bf 76} 031123 (2007).

\bibitem{ManandStan} 
 R. Mantegna and H. E. Stanley, 
 An Introduction to Econophysics Correlations and Complexity in Finance, Cambridge University Press (2000).
\bibitem{laloux}
 L. Laloux, P. Cizeau, J.-P. Bouchaud and M. Potters
 Phys. Rev. Lett. {\bf83} 1467 (1999).
\bibitem{Plerou1}
 V. Plerou, P. Gopikrishnan, B. Rosenow, L. A. N. Amaral, and H. E. Stanley,
 Phys. Rev. Lett. {\bf83} 1471 (1999).
\bibitem{Plerou2}
 V. Plerou, P. Gopikrishnan, B. Rosenow, L. A. N. Amaral, T. Guhr, and H. E. Stanley,
 Phys. Rev. E {\bf65} 066126 (2002).
\bibitem{Utsugi}
A. Utsugi, K. Ino, and M. Oshikawa
 Phys. Rev. E {\bf70} 026110 (2004).

\bibitem{Wang}
D. Wang, B. Podobnik, D. Horvatic, I. Grosse, and H. E. Stanley,
Phys. Rev. E {\bf83} 046121 (2011).
\bibitem{metric}
The strict mathematical definition of a metric actually requires that the metric satisfy a fourth axiom, that of positive
definiteness (non-negativity), which both the Mantegna-Sornette metric and ours satisfy trivially.  See: A.~V. Arkhangel'skii, and L.~S. Pontryagin, {\it General Topology I: Basic Concepts and Constructions Dimension Theory, Encyclopaedia of Mathematical Sciences}, (Springer, 1990).
 \bibitem{Norden}
 L.Norden and M.Weber  European Financial Management {\bf15} 529 (2009).
\bibitem{Kwan}
S. H. Kwan   Journal of Financial Economics {\bf40} 63 (1996).
\bibitem{Hammoudeh}
S. Hammoudeh and H. Li 
Journal of Economics and Business {\bf57} 1 (2005).
\bibitem{businessweek}
http://www.businessweek.com/ap/2012-05/D9V3TCDG0.htm
\bibitem{feller}
W. Feller, 
An Introduction to Probability Theory and Its Applications, Wiley 1997 .
\bibitem{Samuelson} 
P. Samuelson (1965). Industrial Management Review {\bf 6}: 41.
\bibitem{Lin}
W.-L. Lin, R. F. Engle, and T. Ito, 
The Review of Financial Studies {\bf 7} 507 (1994)
\bibitem{Lo}
A. W. Lo, and A. C. Mackinlay,
Journal or Econometrics  {\bf45} 181 (1990)
\bibitem{Bouchaud}
J-P Bouchaud, A. Matacz, and M. Potters, 
Phys. Rev. Lett. {\bf 87}, 228701 (2001)
 \bibitem{Perello} 
J. Perello, and J. Masoliver 
Phys. Rev. E {\bf 67}, 037102 (2003)
\bibitem{Cleveland}
W. S. Cleveland, The American Statistician {\bf 35} 54 (1981)
\bibitem{Graham}  
R.~L. Graham,and P. Hell, Annals of the History of Computing {\bf 7}  43 (1985)
\bibitem{Stavins}
S. R. Stavins,
Handbook of Environmental Economics, Elsevier, Edition 1 Volume 1 (2001)

\end{thebibliography}
\end{document}